\documentclass[11pt]{article}

\usepackage[margin=1in]{geometry}
\usepackage{amsmath,amssymb,amsthm,mathtools,booktabs}
\usepackage{algorithm}
\usepackage{algpseudocode}
\usepackage{booktabs}
\usepackage{hyperref}
\usepackage[nameinlink,noabbrev]{cleveref}
\usepackage[T1]{fontenc}
\usepackage[utf8]{inputenc}
\usepackage{lmodern}
\usepackage{xcolor}
\usepackage{listings}
\usepackage{tikz}
\usetikzlibrary{arrows.meta,positioning,calc}
\usepackage{forest}
\usepackage{threeparttable}

\usepackage[
  backend=biber,
  style=authoryear,
  maxcitenames=2,
  maxbibnames=99,
  doi=false,
  isbn=false,
  url=false
]{biblatex}
\definecolor{pyblue}{RGB}{0,0,180}
\definecolor{pygreen}{RGB}{0,120,0}
\definecolor{pygray}{RGB}{105,105,105}
\definecolor{pybg}{RGB}{248,248,248}
\definecolor{pystring}{RGB}{140,0,140}

\lstdefinestyle{pythonstyle}{
  language=Python,
  backgroundcolor=\color{pybg},
  basicstyle=\ttfamily\small,
  keywordstyle=\color{pyblue}\bfseries,
  commentstyle=\color{pygreen},
  stringstyle=\color{pystring},
  numbers=left,
  numberstyle=\tiny\color{pygray},
  frame=single,
  rulecolor=\color{black},
  breaklines=true,
  breakatwhitespace=true,
  showstringspaces=false,
  tabsize=4,
  upquote=true,
  columns=fullflexible,
  keepspaces=true,
  inputencoding=utf8,
  literate=
    {ω}{{$\omega$}}1
    {τ}{{$\tau$}}1
    {ρ}{{$\rho$}}1
    {θ}{{$\theta$}}1
}
\theoremstyle{definition}
\newtheorem{definition}{Definition}

\theoremstyle{plain}
\newtheorem{proposition}{Proposition}

\newcommand{\ceil}[1]{\left\lceil #1 \right\rceil}
\newcommand{\tauK}{\tau}

\title{Recursive Copula Aggregation for Market and Credit Portfolios}
\author{
Luisa Tibiletti\\
Department of Management, University of Torino\\
Corso Unione Sovietica 218 bis, 10134 Torino, Italy\\
Email: luisa.tibiletti@unito.it
\and
Simone Farinelli\\
Core Dynamics GmbH\\
Setzistrasse 7\\
8272 Ermatingen, Switzerland\\
Email: simone@coredynamics.ch
\and
Eric Dal Moro\\
Signal Iduna Reinsurance\\
Bundesplatz 1\\
6300 Zug, Switzerland\\
Email: eric\_dal\_moro@yahoo.com
}
\date{\today}

\begin{document}
\maketitle

\begin{abstract}
Risk aggregation is a central problem in financial risk management and regulatory capital assessment.
While copula-based approaches are widely used due to their tractability and interpretability,
standard implementations rely on flat dependence structures that may fail to capture hierarchical
interactions across risk components.

This paper introduces a hierarchical copula framework for risk aggregation on trees, in which
dependence is specified locally at each aggregation node and propagated through recursive
rank-based reordering. The approach preserves marginal distributions while enabling flexible and
economically interpretable modeling of dependence across multiple aggregation levels.

Copula parameters are calibrated from empirical data using Kendall’s tau, and the methodology is
illustrated using market and credit risk proxies. Robustness is assessed across alternative credit
segments, including investment-grade and high-yield indices. Numerical results show that dependence
strength increases under stress and varies systematically across credit quality, with important
implications for risk measurement.

The proposed framework combines the transparency of copula-based aggregation with the structural
richness of hierarchical models, providing a practical and scalable tool for enterprise risk
management.
This contributes to bridging bottom-up and top-down approaches to risk aggregation.

\vspace{0.5em}


\vspace{0.5em}

\noindent\textbf{Key Messages:}




\begin{itemize}
\item Recursive hierarchical copulas aggregate market and credit portfolios.

\item Empirical dependence calibration is combined with sample reordering.

\item The framework computes VaR and Expected Shortfall transparently.

\item Tail dependence significantly affects hierarchical risk aggregation.
\end{itemize}
\end{abstract}
\section{Introduction}

\paragraph{Risk aggregation in financial practice.}

Risk aggregation is a central problem in the financial industry, arising in
applications ranging from economic capital calculation to regulatory capital
assessment and portfolio risk management. Two broad modeling approaches have
traditionally been employed. The first is the bottom-up construction of stochastic scenario generators,
where joint dynamics of risk factors are specified through multivariate
time series models and simulated to obtain the distribution of portfolio
losses, see for instance \textcite{schoeftner2010aggregation} and
\textcite{farinelli2016framework}. The second is a top-down approach based on
copula functions, where marginal risk distributions are modeled separately and
their dependence structure is specified via a copula, allowing for a direct
construction of the joint loss distribution.

From a regulatory perspective, the Basel framework---including
\textit{Basel III} and its recent revisions often referred to as
\textit{Basel IV}---does not prescribe a specific methodology for dependence
modeling or risk aggregation, but rather requires institutions to adopt
internally consistent and well-justified models for capital adequacy
assessment \parencite{basel2011,basel2019}.
This leaves significant flexibility in the choice between bottom-up and
top-down approaches.

In practice, stochastic scenario generators offer a detailed and flexible
representation of risk factor dynamics, but may become complex, computationally
intensive, and difficult to calibrate consistently across heterogeneous risk
sources. Copula-based approaches, by contrast, provide a parsimonious and
transparent framework for combining marginal risk models, and are therefore
widely used in top-down risk aggregation. While classical copula models---in
particular Gaussian copulas---are known to exhibit limitations in capturing
tail dependence and extreme co-movements, they remain attractive due to their
tractability, robustness, and interpretability.

\paragraph{Relation to the literature.}

Copula-based dependence modeling has become a standard tool in financial
risk management since the seminal contribution of \textcite{li2000copula},
with subsequent developments emphasizing flexibility and robustness in
capturing nonlinear dependence structures. Early contributions such as
\textcite{embrechts2001modelling, embrechts2002} established the role of
copulas in risk aggregation, while \textcite{patton2006modelling} introduced
time-varying and asymmetric dependence structures. For a comprehensive
treatment of copula-based dependence modeling, see \textcite{joe2014dependence}.

A central challenge in risk aggregation is the construction of flexible yet
tractable high-dimensional dependence models. Dynamic copula specifications
based on score-driven models \parencite{creal2013dynamic, oh2017dynamiccopula}
and factor copula constructions \parencite{krupskii2016factorcopula} allow for
time variation and scalability. Vine copula approaches
\parencite{aas2009paircopula, brechmann2013vine, nagler2021vine} provide a
highly flexible framework for high-dimensional dependence structures, but may
become complex in large-scale applications and are not always aligned with
economically meaningful aggregation hierarchies. From a practical perspective,
comparisons between standard and advanced aggregation models highlight the
trade-off between flexibility and tractability \parencite{eling2020risk}.

In parallel, a growing strand of literature has emphasized structured and
hierarchical dependence models for risk aggregation. Early contributions such
as \textcite{arbenz2012hierarchical} develop sample reordering techniques for
hierarchical aggregation, while \textcite{cossette2017hierarchical} and
\textcite{chaoubi2021hierarchical} propose hierarchical Archimedean copula
constructions. More recent work considers the aggregation of dependent risks
using copula representations \parencite{navarro2022dependent} and highlights
the benefits of combining different copula families to improve the modeling of
aggregate loss distributions and reduce model risk \parencite{dias2025copula}.

A key insight from the empirical literature is the importance of tail
dependence in accurately capturing joint extreme events. Empirical studies
such as \textcite{dias2020copula} and \textcite{lopez2022tail} document a
significant increase in dependence during periods of market stress,
highlighting limitations of Gaussian copula models for risk management
applications. Bayesian approaches further account for parameter uncertainty
and model risk in dependence modeling
\parencite{bernardi2017bayesiancopula, klein2025bayesian}.
\paragraph{Contribution.}

While hierarchical copula constructions have been investigated in the
literature, their practical implementation for risk aggregation remains
relatively limited. Existing approaches often focus either on highly flexible
but complex dependence structures, such as vine copulas, or on global
parametric specifications that do not align naturally with the hierarchical
organization of risks in financial institutions.

This paper proposes a hierarchical copula framework for risk aggregation on
trees that emphasizes operational tractability and economic interpretability.
Dependence is specified locally at each aggregation node and propagated
through the hierarchy via a recursive rank-based reordering procedure. This
construction preserves marginal distributions while avoiding high-dimensional
likelihood estimation, making it particularly suitable for large-scale
aggregation problems.

In contrast to flat copula models, the proposed approach captures heterogeneous
dependence across aggregation levels, and in contrast to vine copulas, it
retains a structure that directly reflects the organization of market and
credit risks. The methodology is implemented in a simulation-based framework
that is easily reproducible and compatible with existing enterprise risk
systems.

Empirically, the model is calibrated using market and credit data and evaluated
across multiple credit segments, including investment-grade and high-yield
indices. The results show a systematic increase in dependence under stress and
highlight the impact of dependence specification on risk measures such as
Value-at-Risk and Expected Shortfall. A robustness analysis across alternative
credit proxies confirms the stability of the proposed framework.

Overall, the paper contributes a practically implementable and structurally
transparent approach to dependence modeling that bridges the gap between
bottom-up scenario generation and top-down copula-based aggregation.
\section{Hierarchical aggregation in a tree}

Copulas separate marginal modeling from dependence modeling \parencite{sklar1959, nelsen2006, mcneil2015}.
Hierarchical copula constructions introduce dependence at multiple aggregation levels, aligning with
typical risk management structures.

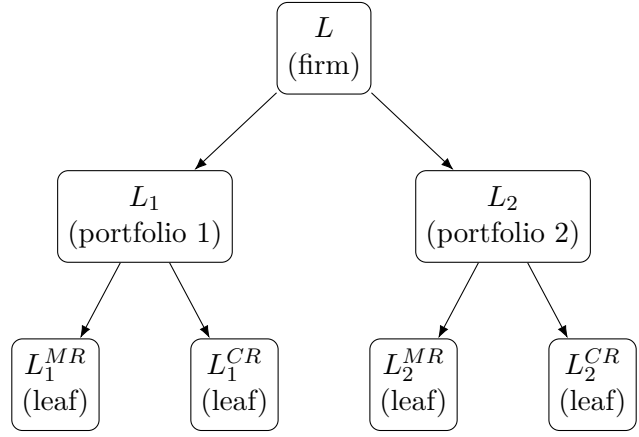
\begin{figure}[h]
\centering
\begin{forest}
for tree={
  draw,
  rounded corners,
  align=center,
  s sep=12mm,
  l sep=10mm,
  edge={-Latex}
}
[$L$\\(firm)
  [$L_1$\\(portfolio 1)
    [$L^{MR}_1$\\(leaf)]
    [$L^{CR}_1$\\(leaf)]
  ]
  [$L_2$\\(portfolio 2)
    [$L^{MR}_2$\\(leaf)]
    [$L^{CR}_2$\\(leaf)]
  ]
]
\end{forest}
\caption{Two-level example: leaf marginals are simulated; dependence is imposed via copulas at internal nodes. MR and CR denote market risk and credit risk, respectively.}
\label{fig:tree}
\end{figure}

\section{Problem statement}

Let the aggregation structure be a rooted tree.
Leaves correspond to marginal loss simulations; internal nodes correspond to aggregated losses defined
as sums of their children.

\begin{definition}[Inputs]
We assume:
\begin{itemize}
\item \textbf{Leaf simulations:} for each leaf variable, a sample of size $n$ is given.
\item \textbf{Node copulas:} for each internal node with $k$ children, a $k$-dimensional copula $C$ is specified to encode dependence among the $k$ child aggregates.
\end{itemize}
\end{definition}

\noindent\textbf{Goal.} Construct a joint simulation over all leaves such that:
(i) leaf marginals are preserved, and
(ii) at every internal node, the child aggregates exhibit dependence consistent with the specified copula.

\subsection{Economic interpretation and loss decomposition}

We interpret the modeled quantity as portfolio loss over a fixed horizon,
defined as the negative of profit and loss:
\[
L = -\mathrm{P\&L}.
\]

To obtain a consistent economic representation, we decompose the loss of
portfolio component $i$ as
\[
L_i = -C_i + L_i^{MR} + L_i^{CR},
\]
where $C_i$ denotes deterministic carry associated with contractual cash
flows, $L_i^{MR}$ represents the market risk component, and $L_i^{CR}$ the
credit risk component.

The carry term $C_i$ is associated with credit exposures and reflects
contractual income such as coupon or interest payments. By contrast, in the
market component, carry effects are already embedded in mark-to-market
valuation and are therefore included in $L_i^{MR}$.

The market component $L_i^{MR}$ captures unexpected mark-to-market
fluctuations and is therefore two-sided:
\[
L_i^{MR} \in \mathbb{R}.
\]
It reflects valuation changes driven by movements in market factors such as
interest rates and credit spreads in the absence of default events.

The credit component $L_i^{CR}$ captures losses arising from default events.
Its sign depends on the nature of the exposure. For portfolios with
non-negative exposures, such as loans or mortgages held to maturity, credit
losses are one-sided:
\[
L_i^{CR} \ge 0.
\]
However, in trading portfolios where exposures may be negative (for instance,
due to credit protection purchased via credit default swaps), default events
may generate gains, and the credit component becomes two-sided:
\[
L_i^{CR} \in \mathbb{R}.
\]

The decomposition therefore accommodates both banking book and trading book
interpretations, depending on the sign of the underlying exposure.
This distinction is important for applications, as regulatory and economic
capital models often treat banking and trading book exposures differently.

Since the carry term $C_i$ is deterministic, it can be removed without loss
of generality by defining the shifted loss variable
\[
\tilde{L}_i = L_i + C_i,
\]
which yields
\[
\tilde{L}_i = L_i^{MR} + L_i^{CR}.
\]
This transformation preserves the dependence structure and does not affect
the copula-based aggregation procedure.

\section{Node aggregation via recursive rank reordering}
The algorithmic construction builds on ideas developed in unpublished
work by Robert Schöftner and Daniel Straumann.
At one internal node with children $i=1,\dots,k$, the algorithm performs a \emph{node\_aggregation} step:

\begin{enumerate}
\item Compute each child aggregate sample $L_i(\omega)$ (sum of all leaves in child $i$'s subtree).
\item Let $\pi_i$ be the permutation that sorts $\{L_i(\omega)\}_{\omega=1}^n$ increasingly.
\item Simulate copula uniforms $(U_1(\omega),\dots,U_k(\omega))\sim C$ for $\omega=1,\dots,n$.
\item Define an index map
\[
j_i(\omega) := \pi_i\!\left(\ceil{nU_i(\omega)}\right).
\]
\item Apply the mapping $\omega\mapsto j_i(\omega)$ to \emph{all leaf samples in child $i$'s subtree}.
\end{enumerate}

Applying this step bottom-up yields a joint simulation consistent with all node copulas while preserving
already-constructed dependence inside each subtree.

\begin{figure}[h]
\centering
\begin{tikzpicture}[
  box/.style={draw, rounded corners, align=center, minimum width=40mm, minimum height=10mm},
  arrow/.style={-Latex, thick},
  lab/.style={font=\small}
]
\node[box] (agg) {Compute child\\aggregate samples\\$L_i(\omega)$};
\node[box, right=18mm of agg] (sort) {Sort aggregates\\get permutations\\$\pi_i$};
\node[box, below=12mm of $(agg)!0.5!(sort)$] (cop) {Sample uniforms\\$(U_1(\omega),\dots,U_k(\omega))\sim C$};
\node[box, below=12mm of cop] (map) {Index map\\$j_i(\omega)=\pi_i(\ceil{nU_i(\omega)})$\\apply to entire subtree};

\draw[arrow] (agg) -- (sort);
\draw[arrow] (agg) -- (cop);
\draw[arrow] (sort) -- (cop);
\draw[arrow] (cop) -- (map);

\node[lab, below=2mm of map] {A single permutation per child is applied to every leaf in that child subtree.};
\end{tikzpicture}
\caption{One node\_aggregation step: copula uniforms choose ranks; ranks map into sorted aggregate samples; induced permutations reorder entire subtrees.}
\label{fig:nodeagg}
\end{figure}

\begin{algorithm}[h]
\caption{Hierarchical copula simulation by recursive node\_aggregation}
\label{alg:hier}
\begin{algorithmic}[1]
\Require Tree with leaf simulations; copula $C$ at each internal node
\State Recurse bottom-up through the tree
\For{each internal node}
  \State Compute each child aggregate sample $L_i(\omega)$
  \State Compute sorting permutation $\pi_i$ for each child aggregate
  \State Sample copula uniforms $(U_1(\omega),\dots,U_k(\omega)) \sim C$
  \State Set $j_i(\omega)=\pi_i(\ceil{nU_i(\omega)})$
  \State Apply mapping $\omega\mapsto j_i(\omega)$ to all leaves in child $i$ subtree
\EndFor
\State \Return joint simulation of all leaves and their aggregates
\end{algorithmic}
\end{algorithm}

Because the carry term $C_i$
is deterministic, shifting the loss variable perfectly preserves the rank-ordering of the samples, ensuring that the copula uniforms $U_k(\omega)$ and the subsequent recursive rank-reordering steps remain completely unaffected.

Although marginal samples are initially generated independently at the leaf
level, the recursive copula-based reordering procedure induces a joint
dependence structure across all components. In particular, dependence between
market risk components across different sub-portfolios is fully captured
through the hierarchical copula construction, and no conditional independence
between them is assumed.

The recursive reordering procedure can furthermore be interpreted as a
constructive implementation of a hierarchical copula, ensuring consistency
with Sklar's theorem at each aggregation level.

\section{Calibration note: Kendall's \texorpdfstring{$\tau$}{tau} to parameters}

Dependence is often specified by Kendall's tau $\tauK$ \parencite{nelsen2006, mcneil2015}.
For two-dimensional copulas, common parameter mappings include:

\begin{proposition}[Gaussian copula]
For the bivariate Gaussian copula with correlation parameter $\rho\in(-1,1)$,
\[
\tauK=\frac{2}{\pi}\arcsin(\rho),
\qquad
\rho=\sin\!\left(\frac{\pi}{2}\tauK\right).
\]
\end{proposition}

\begin{proposition}[Mirrored (survival) Clayton copula]
For the bivariate mirrored Clayton copula with parameter $\theta>0$,
\[
\tauK=\frac{\theta}{\theta+2},
\qquad
\theta=\frac{2\tauK}{1-\tauK}.
\]
\end{proposition}

\section{Empirical calibration workflow}

The empirical analysis is based on daily observations of market and credit
indices from April 2016 onward. Two time windows are considered: the full
sample and a crisis subwindow corresponding to the COVID-19 market
dislocation. The data are aligned on common dates, and dependence is estimated
using Kendall's tau.

A practical calibration is built from one market proxy and one credit proxy observed on the same calendar.
For a public and easily reproducible example, we use the daily S\&P~500 index level as the market series
and daily ICE BofA option-adjusted spread series from FRED as the credit series.
The S\&P~500 is provided in FRED as series \texttt{SP500}; candidate credit proxies include the
BBB corporate OAS series \texttt{BAMLC0A4CBBB}, the broad investment-grade corporate OAS series
\texttt{BAMLC0A0CM}, and the high-yield OAS series \texttt{BAMLH0A0HYM2}
\parencite{fred_sp500, fred_bbb_oas, fred_ig_oas, fred_hy_oas}.

Let $S_t$ denote the market index level and let $\mathrm{OAS}_t$ denote the credit spread series.
We define loss-oriented daily inputs by
\[
X_t = -\log\!\left(\frac{S_t}{S_{t-1}}\right),
\qquad
Y_t = \mathrm{OAS}_t - \mathrm{OAS}_{t-1}.
\]
The sign convention for $X_t$ converts large negative equity returns into positive loss observations,
making both coordinates loss-like.
After aligning the time series and removing missing values, the empirical Kendall coefficient
$\widehat{\tau}$ is computed from the paired sample $\{(X_t,Y_t)\}$ and then mapped into the chosen
copula parameter using the formulas above.

\begin{table}[h]
\centering
\caption{Public series used for calibration and robustness analysis.}
\label{tab:series}
\renewcommand{\arraystretch}{1.15}
\begin{tabular}{lll}
\toprule
Role & Series ID & Interpretation \\
\midrule
Market proxy & \texttt{SP500} & S\&P~500 index level \\
Main credit proxy & \texttt{BAMLC0A4CBBB} & BBB corporate OAS \\
Robustness proxy & \texttt{BAMLC0A0CM} & Investment-grade corporate OAS \\
Robustness proxy & \texttt{BAMLH0A0HYM2} & High-yield corporate OAS \\
\bottomrule
\end{tabular}
\end{table}

\begin{table}[h]
\centering
\begin{threeparttable}
\caption{Dependence calibration based on Kendall's tau. }
\label{tab:calibration}
\renewcommand{\arraystretch}{1.2}
\begin{tabular}{llrrrr}
\toprule
Credit proxy & Window & $N$ & $\tau$ & $\rho_{\mathrm{Gauss}}$ & $\theta_{\mathrm{Clayton}}$ \\
\midrule
\multicolumn{6}{l}{\textit{BBB}} \\
BBB & Full sample & 2514 & 0.2848 & 0.4327 & 0.7966 \\
BBB & COVID & 50 & 0.4035 & 0.5922 & 1.3527 \\
\midrule
\multicolumn{6}{l}{\textit{IG}} \\
IG & Full sample & 2513 & 0.2732 & 0.4161 & 0.7519 \\
IG & COVID & 50 & 0.3926 & 0.5783 & 1.2927 \\
\midrule
\multicolumn{6}{l}{\textit{HY}} \\
HY & Full sample & 2514 & 0.3741 & 0.5544 & 1.1954 \\
HY & COVID & 50 & 0.5166 & 0.7252 & 2.1369 \\
\bottomrule
\end{tabular}

\begin{tablenotes}
\footnotesize
\item IG and HY denote investment-grade and high-yield credit segments, respectively.
\end{tablenotes}
\end{threeparttable}
\end{table}

The results are stable across credit segments. Dependence increases systematically from investment-grade
to high-yield credit, reflecting the higher sensitivity of lower-quality credit to market conditions.
During the COVID-19 crisis, Kendall's $\tau$ increases markedly for all segments, indicating stronger
comovement under stress. The mirrored Clayton parameter exhibits a particularly strong increase,
highlighting the amplification of lower-tail dependence during crisis periods.

\section{General implementation for arbitrary trees}

The original two-level example is retained for exposition, but the implementation is now written for
arbitrary rooted trees.
Each internal node stores:
(i) a list of children,
(ii) a copula sampler,
and optionally
(iii) metadata such as the copula family, the calibrated Kendall's tau, and the fitted parameter.
The same bottom-up recursive reordering algorithm then applies regardless of depth or local branching number.
Only the local copula dimension changes from node to node.

This generalization separates the mathematical construction from specific
illustrative examples. While the exposition focuses on a two-level aggregation
structure as in Figure \ref{fig:tree}, the framework extends naturally to deeper
hierarchies, such as desk--portfolio--business-line--group aggregation trees,
without requiring conceptual modifications.

The numerical implementation is based on a recursive Python prototype
developed for reproducibility. The implementation code is publicly available
in a dedicated GitHub repository.

\section{Numerical illustration}

To keep the numerical illustration transparent, we retain the two-level tree of Figure \ref{fig:tree} and use
illustrative dependence inputs at the three internal nodes:
\[
\tau_{p_1}=0.54,\qquad
\tau_{p_2}=\frac{1}{3},\qquad
\tau_{\mathrm{root}}=\frac{1}{3}.
\]
For the Gaussian specification this gives
\[
\rho_{p_1}=\sin\!\left(\frac{\pi}{2}\cdot 0.54\right)\approx 0.7501,
\qquad
\rho_{p_2}=\rho_{\mathrm{root}}=\sin\!\left(\frac{\pi}{6}\right)=0.5,
\]
while for the mirrored Clayton specification it gives
\[
\theta_{p_1}=\frac{2\cdot 0.54}{1-0.54}\approx 2.3478,
\qquad
\theta_{p_2}=\theta_{\mathrm{root}}=\frac{2/3}{2/3}=1.
\]

\begin{table}[h]
\centering
\caption{Illustrative node-level dependence inputs used in the numerical example.}
\label{tab:nodeparams}
\renewcommand{\arraystretch}{1.15}
\begin{tabular}{lrrr}
\toprule
Node & Kendall's $\tau$ & Gaussian parameter $\rho$ & Mirrored Clayton parameter $\theta$ \\
\midrule
$p_1$ & 0.5400 & 0.7501 & 2.3478 \\
$p_2$ & 0.3333 & 0.5000 & 1.0000 \\
$\mathrm{root}$ & 0.3333 & 0.5000 & 1.0000 \\
\bottomrule
\end{tabular}
\end{table}

Aggregate-loss summary statistics are reported in \cref{tab:riskresults}. These numbers are reproducible from the bundled code
and serve as a stable baseline for later empirical calibration.

\begin{table}[h]
\centering
\caption{Aggregate-loss summary statistics for the illustrative two-level example.}
\label{tab:riskresults}
\renewcommand{\arraystretch}{1.15}
\begin{tabular}{lrrrr}
\toprule
Copula family & Mean & Std.\ dev. & VaR 95\% & ES 95\% \\
\midrule
Gaussian & 36.2453 & 2.9314 & 41.3056 & 44.2413 \\
Mirrored Clayton & 36.2525 & 3.2129 & 42.0276 & 45.9763 \\
\bottomrule
\end{tabular}
\end{table}

The marginal distributions at the leaf level are specified independently
of the copula structure. In the present implementation, simple parametric
distributions are used for illustration, while the dependence structure is
entirely captured by the copula. Details on the marginal specifications are
provided in Appendix~\ref{appA}.

\section{Risk Implications of Dependence Structure}

We now evaluate the impact of the calibrated dependence structures on
portfolio risk measures using Monte Carlo simulation within the
hierarchical copula framework.

\begin{table}[h]
\centering
\caption{Risk measures under alternative copula specifications.}
\label{tab:var_es}
\renewcommand{\arraystretch}{1.2}
\begin{tabular}{llcc}
\toprule
Credit proxy & Scenario & VaR 95\% & ES 95\% \\
\midrule
\multicolumn{4}{l}{\textit{BBB}} \\
BBB & Gaussian (Full) & 41.0407 & 44.1091 \\
BBB & Gaussian (COVID) & 41.3350 & 44.6173 \\
BBB & Clayton (Full) & 41.6532 & 46.0990 \\
BBB & Clayton (COVID) & 42.1060 & 46.6626 \\
\midrule
\multicolumn{4}{l}{\textit{IG}} \\
IG & Gaussian (Full) & 40.9427 & 44.3582 \\
IG & Gaussian (COVID) & 41.4503 & 44.7726 \\
IG & Clayton (Full) & 41.2999 & 45.4799 \\
IG & Clayton (COVID) & 41.9913 & 46.5444 \\
\midrule
\multicolumn{4}{l}{\textit{HY}} \\
HY & Gaussian (Full) & 41.3466 & 44.5190 \\
HY & Gaussian (COVID) & 41.7954 & 45.2575 \\
HY & Clayton (Full) & 41.8431 & 45.5463 \\
HY & Clayton (COVID) & 42.3191 & 47.1069 \\
\bottomrule
\end{tabular}
\end{table}

The results highlight the importance of tail dependence for risk
measurement. While Gaussian and mirrored Clayton copulas yield similar
Value-at-Risk estimates, the Expected Shortfall is systematically higher
under the mirrored Clayton specification.

This effect is particularly pronounced during the COVID crisis and for
high-yield credit, reflecting the role of lower-tail dependence in
amplifying joint extreme losses.

These findings suggest that Gaussian copulas may underestimate tail risk
in stressed market conditions, whereas the mirrored Clayton copula
provides a more conservative and structurally consistent assessment.

\subsection{Why hierarchical copulas outperform flat copulas}

A flat copula imposes a single dependence structure on all risk factors at
once. In the present setting this would force market and credit losses
from heterogeneous portfolio segments to share one common copula
parameterization, thereby ignoring the layered architecture of risk
aggregation.

By contrast, a hierarchical copula model assigns dependence locally at
each internal node of the aggregation tree. This has three advantages.
First, it allows within-portfolio dependence between market and credit
losses to differ from cross-portfolio dependence at the firm level.
Second, it preserves economic interpretability, since each copula is
attached to a well-defined aggregation step. Third, it permits stressed
dependence to propagate through the hierarchy in a structurally coherent
way, which is especially important for tail-sensitive measures such as
Expected Shortfall.

The robustness results confirm this point. Dependence rises from
investment-grade to high-yield credit and increases further in the COVID
crisis window. A hierarchical specification can incorporate these effects
naturally at the relevant portfolio levels, whereas a flat copula would
compress them into a single global parameter and thereby lose information
about where dependence is generated and how it amplifies risk across the
tree.

\section{Conclusion}
This paper develops a hierarchical copula framework that enhances classical
top-down risk aggregation by incorporating structural dependence across levels. More exactly, it 
 proposes a hierarchical copula framework for the aggregation of
market and credit risks, motivated by the need for tractable and interpretable
models of dependence in financial risk management. In contrast to flat copula
approaches, the hierarchical structure reflects the underlying organization of
risk factors and allows dependence to be specified locally at each aggregation
level.

The empirical analysis, based on Kendall's tau calibration using market and
credit indices, highlights the importance of tail dependence, particularly
during periods of market stress such as the COVID-19 crisis. Simulation results
show that tail-dependent copula specifications lead to materially higher
risk measures, as reflected in Value-at-Risk and Expected Shortfall estimates.

Robustness checks across different credit proxies confirm the stability of the
main findings and illustrate the flexibility of the proposed framework in
capturing heterogeneous dependence structures.

The results suggest that hierarchical copula models provide a compelling
alternative to both purely bottom-up scenario generators and standard flat
copula approaches, combining interpretability with improved representation of
tail risk. Future research may explore extensions to higher-dimensional
settings, dynamic dependence structures, and integration with machine learning
techniques for enhanced calibration.

\section*{Competing interests}

\section*{Acknowledgment on AI usage}
Parts of the exposition and code structuring benefited from AI-assisted tools (Wolfram 5.1 and Chat GPT Plus),
which were used from February 2026 till April 2026 to improve clarity and organization. Parts of the bibliography was researched using AnswerThis in April 2026. All modeling choices,
results, and interpretations remain the responsibility of the authors.

\section*{Corresponding author contact details}
Luisa Tibiletti, Department of Management, University of Torino, Italy. Email: luisa.tibiletti@unito.it

\printbibliography

\appendix

\section{Appendix: Marginal models and dependence calibration}\label{appA}

The copula-based aggregation framework separates marginal modeling from
dependence specification. In the numerical implementation, these two
components are treated distinctly.

At the leaf level, marginal distributions are specified parametrically for
transparency and reproducibility. Market risk components are modeled using
lognormal distributions, reflecting the multiplicative nature of price
dynamics and the resulting asymmetry in loss profiles. Credit risk components
are modeled using shifted Pareto distributions, capturing the heavy-tailed
behavior associated with default-driven losses.

The dependence structure is calibrated independently using empirical data.
Specifically, market and credit time series obtained from publicly available
sources are aligned on common observation dates and transformed into
loss-oriented quantities. Kendall's tau is then estimated from the empirical
pairs and mapped into the parameters of the selected copula families.

The simulation of aggregate losses combines these two elements: parametric
marginal distributions at the leaf level and empirically calibrated dependence
parameters. This approach preserves the flexibility of the copula framework
while ensuring that dependence is informed by observed data.

It should be emphasized that the choice of marginal distributions is not
restrictive. Alternative specifications, including empirical or heavy-tailed
parametric distributions, can be incorporated without modification of the
aggregation procedure, as the dependence structure is entirely governed by
the copula.
\section{Implementation details and Python code}

The Python code in \texttt{hierarchical\_copula\_engine.py} implements:
\begin{itemize}
\item Gaussian and mirrored Clayton copula sampling
\item calibration helpers from historical data and FRED CSV exports
\item the one-node node\_aggregation step
\item the full recursion for arbitrary rooted trees
\item a two-level numerical illustration and automatic rendering of a LaTeX results table
\end{itemize}
The implementation code supporting the empirical results is publicly
available at:\\
\url{https://github.com/SimoneFarinelli10/hierarchical-copula-aggregation}.

The implementation assumes that input data are stored in a local
\texttt{data/} directory.


\subsection{Reproducibility}

To run the numerical illustration locally:
\begin{lstlisting}
python hierarchical_copula_engine.py
\end{lstlisting}

To calibrate from public data, export the chosen FRED series as CSV files over a common date range
and call \texttt{calibrate\_bivariate\_pair\_from\_fred\_csv} with the corresponding paths.

\end{document}